\documentclass{iau}

\usepackage{graphicx}
\usepackage{verbatim}
\usepackage{hyperref}
\usepackage[textwidth=1.5cm,shadow]{todonotes}	% \todo{}: side notes at the margins; \todo[inline]{}: inline notes; \missingfigure{}; \listoftodos
\usepackage{color, soul}	% Highlighting, use \hl{...}
\usepackage{fontawesome}    % lots of emoji

\newcommand{\orcid}[1]{\href{#1}{\includegraphics[scale=0.035]{figures/orcid.png}}}

\title[IAUS359. AI and black holes]{The first AI simulation of a black hole}

\author[Nemmen, Duarte \& Navarro]   %% give here short author list %%
{Rodrigo Nemmen$^1$,
Roberta Duarte$^1$ \and Jo\~ao P. Navarro$^2$}

\affiliation{$^1$Universidade de S\~ao Paulo, Instituto de Astronomia, Geof\'{\i}sica e Ci\^encias Atmosf\'ericas, Departamento de Astronomia, S\~ao Paulo, SP 05508-090, Brazil \\ Email: {\tt rodrigo.nemmen@iag.usp.br} \\[\affilskip]
$^2$NVIDIA} %\\email: {\tt hoefner@astro.uu.se}}

\pubyear{2020}
\volume{359}  %% insert here IAU Symposium No.
\jname{Galaxy evolution and feedback across different environments}
\editors{T. Storchi-Bergmann, R. Overzier, W. Forman \& R. Riffel, eds.}

\begin{document}

\maketitle

\begin{abstract}
We report the results from our ongoing pilot investigation of the use of deep learning techniques for forecasting the state of turbulent flows onto black holes. Deep neural networks seem to learn well black hole accretion physics and evolve the accretion flow orders of magnitude faster than traditional numerical solvers, while maintaining a reasonable accuracy for a long time.  
\keywords{Black hole physics, astrostatistics, active galactic nuclei}
%% add here a maximum of 10 keywords, to be taken form the file <Keywords.txt>
\end{abstract}

\firstsection % if your document starts with a section,
              % remove some space above using this command.
\section{Introduction}

My presentation was supposed to be about the new constraints on the spin of the supermassive black hole (SMBH) in M87. Here is the bottom line from that work: we have constrained the spin parameter to be $|a_*| > 0.4$ (\cite[Nemmen 2019]{Nemmen2019}). The spin is the second fundamental parameter of black hole (BH) spacetimes. This constraint should set expectations for future estimates of the M87* spin with the Event Horizon Telescope and other observatories.

Instead, I will present some early exciting results from our pilot investigation of artificial intelligence (AI) methods as tools to accelerate numerical simulations of BH accretion flows. Here, we address two inter-related questions: Can we make the models faster while maintaining an accuracy comparable to explicit solvers of the fluid conservation equations? Can deep neural networks learn fluid dynamics?

\section{Deep learning}

Let me begin with the fundamental problem of BH astrophysics: to figure out the function 
\begin{equation}
{\rm AGN}(t) = f(M, a_*, \dot{M})
\end{equation}
which quantifies the complete time-evolution of ``weather'' around SMBHs, where $M$ and $\dot{M}$ are the BH mass and mass accretion rate, respectively. The challenge is that BH weather is a complex process, requiring the solution of  nonlinear, multidimensional partial differential equations which are very time-consuming (e.g. \cite[Porth et al. 2019]{Porth2019}). Ideally, we want those simulations to have a duration much shorter than a typical PhD thesis timescale, so we need to make them as fast as possible.

Here, we are investigating the use of deep learning (DL) techniques for that purpose. DL consists of using deep neural networks inspired by the way the brain works, with a large number of layers and parameters (``neurons'') (\cite[Goodfellow, Bengio \& Courville 2016]{Goodfellow2016}). Deep neural networks are good approximators for empirical functions which are too complex to be have an analytical form (\cite[e.g. Cybenko 1989]{Cybenko1989}). It is not an exaggeration to say that a considerable fraction of AI work today consists of applications---and improvements upon---DL.

In practice, DL algorithms ``learn from experience'': instead of explicitly coding the instructions in the code, one trains the machine by showing a lot of examples and comparing the output of trained algorithm to a test dataset. Then, by tweaking the network architecture and its hyperparameters, one arrives at a trained deep net (\cite[LeCun 2015]{LeCun2015}). DL is leading to several breakthroughs in many fields (e.g. \cite[Mnih 2015]{Mnih_etal15};  \cite[Silver 2016]{Silver_etal16}; \cite[Krizhevsky 2017]{Krizhevsky_etal17}), including astronomy (e.g. \cite[Hausen 2020]{Hausen_etal20};  \cite[Zhang 2019]{Zhang_etal20}). The downsides of DL is that it is data-hungry (needs a lot of data to be effective) and computationally expensive to train. Once trained, however, the algorithm deploys answers very quickly. 

Here, we devise the prediction challenge as a computer vision problem and infer the evolution of the system from the sequence of input data cubes that comprise previous states of an accreting BH. This is a data-driven, equation-free approach in which a model learns to approximate the relevant physics from the training examples alone and not by incorporating a priori knowledge about the equations underlying the processes. This is similar to the work of \cite[Jaeger \& Haas (2004)]{Jaeger_etal04}; \cite[Tompson et al. (2016)]{Tompson_etal16}.

\section{Teaching a machine about BH accretion}

The training data set consists of the hydrodynamical BH accretion simulations performed by \cite[Almeida \& Nemmen (2020)]{Almeida2020}. This is a set of long-duration (durations $> 10^5 GM/c^3$), 2D models designed to explore the winds produced by low-luminosity active galactic nuclei, where a Schwarzschild BH is fed by a hot, geometrically thick accretion flow. The specific model we chose is PNSS3, which has one of the longest durations among the models computed by \cite[Almeida \& Nemmen (2020)]{Almeida2020}. The numerical simulation computes the spacetime distribution of gas densities around the BH which we feed to the DL model.

For the learning algorithm, we use the well-known U-Net convolutional neural network (ConvNet) architecture, which is commonly used to extract information from datasets which involve spatial and temporal coherence (e.g. \cite[Karpathy 2014]{Karpathy_etal14}). For the training, we divide the time series data into 67.5\% training, 12.5\% cross-validation and 20\% testing phases. How well does the trained DL model predict the future state of the density field around the BH?

\section{Preliminary results}

We quantify the performance of the DL approach by comparing a number of indicators with those obtained from the explicit solution to the conservation equations (Duarte, Nemmen \& Navarro, in preparation). Here, we present some early, preliminary results.

\begin{figure}
\includegraphics[width=\linewidth]{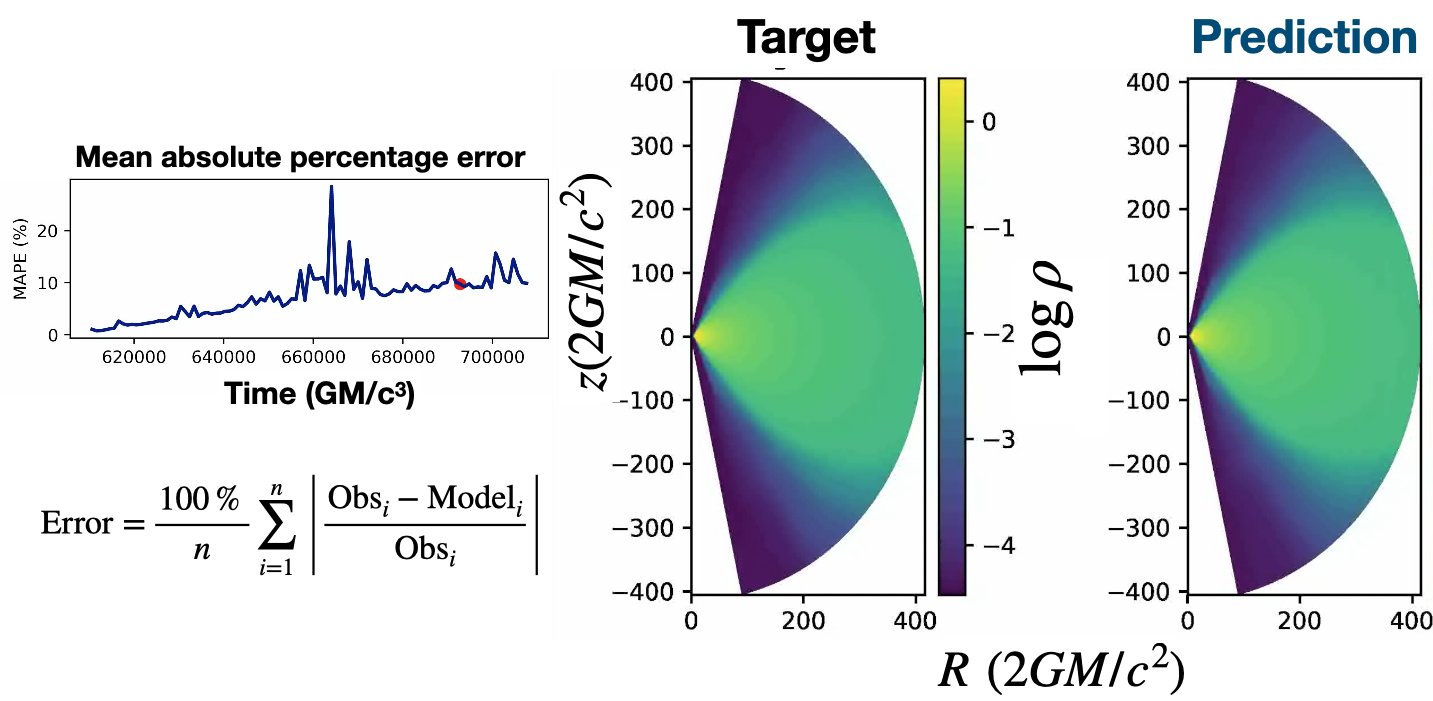}
\caption{Comparison between the deep neural network's ``imagination'' (indicated as \textit{prediction}) and the actual numerical solutions from the fluid conservation laws (the \textit{target}), for a black hole surrounded by a hot accretion flow. Left panel: time evolution of the difference between the prediction and target (i.e. the error of the trained model). The lengths are expressed in terms of the Schwarzschild radius. Time in gravitational units. } 
\label{fig01}
\end{figure}

The left panel of Figure \ref{fig01} displays the density difference between the numerical simulations (i.e. the target) and the trained ConvNet (the prediction) as a function of time. What we mean by error here is defined in the lower left corner of Figure \ref{fig01}. We see that the error gradually builds up over time, as if the predictions drifted from the ground truth solutions of the conservation equations. The middle panel shows the target density map, which results from solving the fluid conservation equations. The right panel corresponds to the inferred deep learning model---the neural network's imagination. 

\begin{figure}
\includegraphics[width=\linewidth]{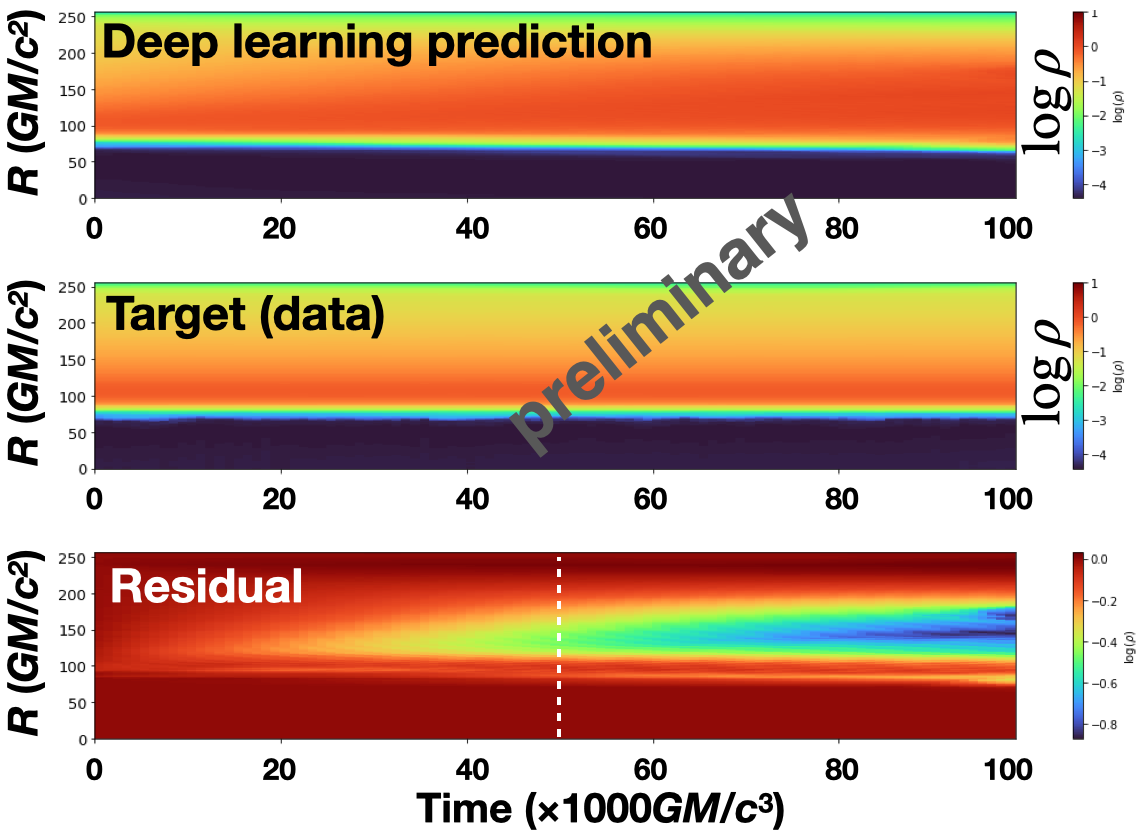}
\caption{Prediction of deep learning model compared with the data from the numerical simulations of BH accretion. The colors map the logarithm of the gas density. The density is averaged over the polar angle. The residuals are the difference between the logarithms. The vertical white line indicates the time at which the deep nets begin to drift away from the ground truth. }
\label{fig02}
\end{figure}

Figure \ref{fig02} summarizes the main result of this presentation. Each panel displays the density marginalized over the polar angle. Think of this as the average density in spherical shells of increasing radii. On a first inspection, we can see that the DL forecast and the data are virtually identical. Only after analyzing the residuals we realize that there are differences between the two. We see that the neural network imagines the future well for a duration of about $5 \times 10^4 GM/c^3 \approx 33 t_{\rm dyn}$, where $t_{\rm dyn}$ is the dynamical time at 100 Schwarzschild radii. Relatively speaking, this is a long time considering the timescales that regulate the system. Only after $33 t_{\rm dyn}$ the neural network begins displaying symptoms of an ``artificial Alzheimer's disease''.  

These results indicate that DL techniques are promising for forecasting the state of multidimensional chaotic systems. While the trained model is performing well---i.e. reproducing the target data accurately---it does so much faster than the original simulation: about \textit{600 times faster}. The DL training was done using two NVIDIA GPUs ($\approx 40$ TFlops, FP32); the hydrodynamic simulation was  performed on a CPU cluster using 200 cores in parallel ($\sim 3$ TFlops, FP64). Even taking into account the different performances of the processors used, the DL model is still vastly faster than the original modeling approach used to generate the training data.

\section{Summary}

This work is a pilot study of the performance of deep learning in forecasting the state of large spatiotemporally chaotic systems comprised by turbulent flows onto black holes. So far, we have obtained promising preliminary results. Not only deep neural networks seem to learn well black hole accretion physics: they evolve the accretion flow $\sim 1000$ times faster than traditional numerical solvers. The future seems bright for black hole numerical investigations, in which science will be less restricted by hardware limitations. 

This contribution is a brief teaser of the full results, which will be reported in Duarte, Nemmen \& Navarro (in preparation). We gratefully acknowledge support by FAPESP (Funda\c{c}\~ao de Amparo \`a Pesquisa do Estado de S\~ao Paulo) under grant 2017/01461-2, CAPES, and the support of NVIDIA Corporation with the donation of the Quadro P6000 GPU used for this research.

\begin{discussion}

\discuss{Weinberger}{From the movie you showed us, it seems that the accretion flow is almost stationary. I do not see much turbulence going on.}

\discuss{Nemmen}{We have chosen for the training data a model from Almeida \& Nemmen (2020) that shows little variability, because this was the longest model we had (more data improves the performance of the training). We are also investigating and generating data with much more variability and comparable durations.}

\discuss{Sanchez}{I heard that there are also some GPU codes that solve fluid dynamics and are faster than CPU solvers. Did you compare the speedups of your deep learning approach with those GPU solvers?}

\discuss{Nemmen}{I only know of one GPU solver for black hole accretion simulations, H-AMR developed by Liska et al. In my understanding, H-AMR is obtaining speedups of about 10x when comparing GPUs and multicore CPUs. Our early results indicate a speedup of about 1000x using a good GPU of the kind that gamers use. So it seems that a DL approach would offer a great cost-benefit. But much remains to be investigated yet.} 

\discuss{Wylezalek}{Can you trust the predictions of an algorithm that does not solve the physical laws? In other words, can you really do physics without doing physics? You don't know what the AI algorithm is really doing, how it arrives at a given answer.}

\discuss{Nemmen}{This is a big issue in AI nowadays: the issue of explainability---explaining how the neural networks arrive at a given output. I do not have an answer for that. But I do know that there is progress in this front by some AI research groups. Using AI for the purpose described in this presentation involves a shift in the underlying philosophy of the numerical simulations. What if it does indeed work well for forecasting while being much faster than current simulations? I think this is a very interesting question.} 

\end{discussion}

\end{document}